\documentclass[aps,prl,twocolumn,groupedaddress,showpacs]{revtex4}
\usepackage{graphicx}
\def\sun{\odot}
\def\gsim{\agt}

\begin{document}

\title{The Dark Matter Profile in the Galactic Center}

\author{Oleg Y. Gnedin}
\affiliation{Space Telescope Science Institute \\ 
  {\tt ognedin@stsci.edu}}

\author{Joel R. Primack}
\affiliation{Physics Department, University of California, Santa Cruz \\
  {\tt joel@scipp.ucsc.edu}}

\date{\today}

\begin{abstract}

We describe a quasi-equilibrium profile of dark matter particles in the inner
parsec of the Galaxy, $\rho_{\rm dm} \propto r^{-3/2}$.  This
``mini-cusp'' profile is caused by scattering with the
dense stellar cluster around the supermassive black hole in Sgr A$^*$
and is independent of the initial conditions.  The implications for
detection of gamma rays from WIMP dark matter annihilation in the Galactic
center are a mild enhancement of the flux, and a characteristic
central feature in the angular distribution which could be detectable
by high-resolution Atmospheric Cherenkov Telescopes.

\end{abstract}

\maketitle

The distribution of dark matter in the very center of our Galaxy is
important for experimental searches for signatures of possible
annihilation of supersymmetric dark matter particles.  Such
weakly interacting massive particles (WIMPs) have
remained favored candidate for the dark matter that represents most of
the mass of the universe ever since it was first proposed that the
lightest supersymmetric partner particle, stable as a result of
R-parity conservation, is the nonbaryonic dark matter \cite{PP:82}.
In current standard supersymmetric theories these are expected to be
neutralinos $\chi$ \cite{Goldberg} which naturally have the required
cosmological density $\Omega_\chi \sim 0.25$ \cite{JKG:96}.

Since dark matter makes a negligible contribution to the dynamical mass
in the central parsec, its distribution cannot be probed directly.
Instead, it should be inferred considering all relevant physical
processes in the Galactic center.  Previous studies have reached
different, often conflicting conclusions based on various initial
assumptions and processes considered.  All previous work assumed that
dark matter particles are collisionless and therefore
conserve their phase-space density.

\begin{itemize}

\item
On scales above $\sim1$ kpc, most dissipationless simulations of
galaxy formation predict a power-law cusp in the dark matter density,
$\rho_{\rm dm} \propto r^{-\gamma}$ with $\gamma = 1-1.5$
\cite{NFW:97,Moore:99,Klypin:01,Power:03}, or perhaps even $\gamma <
1$ \cite{Stoehr:03}.

\item
In the vicinity of the Galactic center, the supermassive black hole
with $M_{\rm bh} = 3.7 \times 10^6 \, M_{\sun}$ dominates the mass in
the inner $r< r_{\rm bh} = 2$ pc \cite{Schoedel:02,Genzel:00,Ghez:98}.
If the central black hole grew adiabatically from a small seed, for
example by accretion of gas, stars, and dark matter
\cite{LBR:71,ZHR:02}, the dark matter cusp would be enhanced and would
form a \textit{spike}, $\rho_{\rm dm} \propto r^{-A}$ with $A =
(9-2\gamma)/(4-\gamma) \approx 2.3-2.4$ \cite{GS:99}.

\item
If instead the black hole appeared instantaneously, being brought in
by mergers of progenitor halos, the enhancement is weaker and the
spike has a slope $A=4/3$ \cite{UZK:01}.  A combination of this and
the previous effect, such that the mergers create a seed black hole that
later grows by accretion, produces an intermediate slope.

\item
Possible mergers of black holes in the centers of the progenitor halos
\cite{MM:01,Volonteri:03} may have the opposite effect on the dark
matter distribution.  Numerical simulations \cite{MMVJ:02} show that
the kinetic heating of particles during the merger may reduce their
density to a very weak power-law, $\rho_{\rm dm} \propto r^{-1/2}$.
However, the simulations do not extend further than about 1 pc towards
the center and thus cannot tell us about the very inner profile.

\item
The same weak cusp, $A = 1/2$, results if the black hole grows away
from the center of the dark matter distribution \cite{ZN:71,UZK:01}.

\end{itemize}

Thus there is a considerable ambiguity in what the inner dark matter
profile could be, all due to the uncertainty in the history of the
central region of our Galaxy.

\bigskip
{\it Scattering off stars sets a quasi-equilibrium profile.---} The above
considerations assumed that the phase-space density of dark matter
particles is conserved.  However, in addition to the supermassive
black hole, the Galactic center harbors a compact cluster of stars,
with the density $\rho_* = 8 \times 10^8 \, M_{\sun}$ pc$^{-3}$
\cite{Genzel:03} in the inner 0.1'' (0.004 pc at the distance of
the Sun of 8 kpc).  These stars frequently scatter dark matter
particles and cause the distribution function to evolve towards an
equilibrium solution.  Both stars and dark matter experience two-body
relaxation.

The idealized problem of a stellar distribution around a massive black
hole in star clusters has been considered in the past (cf. \cite{S:85}
for a review).  Stars driven inward towards the black hole by two-body
relaxation try to reach thermal equilibrium with the stars in the
core, but are unable to do so because of tidal disruption or capture
by the black hole.  Unlike core collapse in self-gravitating star
clusters, however, the density of inner stars does not grow toward
infinity.  A steady-state solution is possible where the energy
released by removal of the most bound stars is transported outward by
diffusion.  Because there is no special scale in the problem, the
quasi-equilibrium distribution function is a power-law of energy,
$f(E) \propto E^p$, and the density is a power-law of radius, $\rho
\propto r^{-3/2-p}$ \cite{BW:76,SL:76}.  The solution is unique and
independent of the initial conditions.

Within the sphere of influence of the central black hole $r_{\rm bh}$,
the distribution functions of both stars and dark matter are
determined by two-body scattering and have the above power-law form.
The scatterers in both cases are stars, but the distribution of dark
matter differs in the exponent $p$ from that of stars because of the
vastly different masses of the two species.  The evolution of the dark
matter distribution $f(E,t)$ in a two-component system of dark matter
particles of mass $m_\chi$ and stars of mass $m_*$ can be described by
a collisional Fokker-Planck equation (first derived in this form in
\cite{M:83}):
\begin{eqnarray*}
  -{\partial q \over \partial E} {\partial f \over \partial t} =
  A {\partial \over \partial E}
  \left[ {m_\chi \over m_*} f
         \int_E^\infty f_* {\partial q_* \over \partial E_*} dE_* 
  \right. \\ \left.
       + {\partial f \over \partial E}
       \left\{ \int_E^\infty f_* q_* dE_* 
      + q \int_{-\infty}^E f_* dE_* \right\}
  \right],
\end{eqnarray*}
where $E = G M_{\rm bh}/r - \frac{1}{2} v^2$ is the binding energy per
unit mass, $q(E) = (2^{3/2}/3) \pi^3 G^3 M_{\rm bh}^3 E^{-3/2}$, $A
\equiv 16 \pi^2 G^2 m_*^2 \ln{\Lambda}$, and $\ln{\Lambda} =
\ln{M_{\rm bh}/m_*} \approx 15$ is the standard Coulomb logarithm.
The equilibrium distribution function of stars is $f_*(E_*,t) \propto
E_*^{1/4}$, i.e. $p=1/4$ \cite{BW:76}.  For dark matter particles,
however, the first term in the square brackets vanishes since the
particle mass is negligible compared to stellar mass.  An equilibrium
solution with no energy flux requires $\partial f/\partial E = 0$, or
$p=0$.  The corresponding density profile is $\rho_{\rm dm} \propto
r^{-3/2}$.

Dark matter particles cannot be tidally disrupted by the black hole,
but they will be captured from the loss cylinder where their angular
momentum per unit mass is less than $J_{\rm cap} = 4 G M_{\rm bh}/c$
(in which case their minimum distance to the black hole is within the
last stable orbit).  This would drive the outward flux of energy that
needs to be balanced by the inward flux of particles, just as in the
stellar case.  As long as the relaxation time is shorter than the
Hubble time, the quasi-equilibrium solution should be gradually
achieved over a large range of energies.  Note that this solution
would be broken at the inner boundary where the distribution function
would smoothly vanish due to the particle loss.  The normalization of
the profile may also slowly evolve (see \cite{Merritt:03} for
discussion).

The observed stellar distribution around the black hole is roughly
consistent with the equilibrium slope: $\rho_*(r) = 1.2 \times 10^6 \,
(r/0.4 \, \mbox{pc})^{-\alpha} \, M_{\sun}$ pc$^{-3}$ with $\alpha =
2.0 \pm 0.1$ at $r > 0.4$ pc and $\alpha = 1.4 \pm 0.1$ at $r < 0.4$
pc \cite{Genzel:03}.  The inner slope of the stellar profile is
somewhat shallower than the predicted $\alpha = \frac{3}{2} + p =
\frac{7}{4}$, but the overall profile is consistent with that of a
relaxed cluster.  The inner profile is well measured down to $r =
0.004$ pc from the center, where the local relaxation time is
\cite{SH:71}:
$$
    t_{\rm rel} = 0.065 {v^3 \over G^2 m_* \rho_* \ln{\Lambda}}
                \approx 2\times 10^9 \, \mbox{yr},
$$
where $v(r) \approx (G M_{\rm bh}/r)^{1/2}$ is the rms velocity of
stars and dark matter, and $m_* \approx 1 \, M_{\sun}$.  The density
increase towards the black hole almost exactly balances the increase
of the particle velocities, and the relaxation rate is roughly
independent of radius.  If the $\alpha = 1.4$ stellar profile
continues all the way to the gravitational radius of the black hole,
$r_g \equiv 2GM_{\rm bh}/c^2 = 4\times 10^{-7}$ pc, the relaxation
time is shorter than the Hubble time everywhere.  

The angular momentum of dark matter particles grows on the average as
$(G M_{\rm bh} r)^{1/2}$ and the radius at which it equals $J_{\rm
cap}$ is $L_{\rm cap} = 8 \, r_g \approx 3\times 10^{-6}$ pc.
However, the extrapolated density at $L_{\rm cap}$ is so high that
dark matter particles can annihilate faster then they are scattered
in.  For a typical value of the annihilation cross section
$\left<\sigma_{\rm ann} v\right> \sim 10^{-26}$ cm$^3$ s$^{-1}$
(c.f. Fig. 1), $\rho_0 = 100 M_\odot$ pc$^{-3}$ (see below), 
and $m_{\chi} \sim 100$ GeV c$^{-2}$, the annihilation
time equals the relaxation time at $L \approx 10^{-5}$ pc.  We
therefore take this value of $L$ as the smallest radius at which we
can trust our dark matter profile.

Note also that stellar-mass black holes resulting from stellar
evolution of ordinary stars would sink towards the center from the
inner 5 pc and would form a tight cluster of their own \cite{MG:00}.
This cluster scatters dark matter particles as well and additionally
contributes to the relaxation.

\bigskip
{\it Implications for dark matter searches.---} The dark matter density
in the central region of the Galaxy is thus given by
$$
\rho_{\rm dm}(r)=\left\{ \begin{array}{ll}
 \rho_0 \left({r/r_{\rm bh}}\right)^{-3/2}  & \qquad L < r 
    \le  r_{\rm bh} \ ,\\ 
 \rho_0 \left({r/r_{\rm bh}}\right)^{-\alpha} & \qquad r_{\rm bh}
    \le r \ , 
\end{array} \right.
$$
where $L = 10^{-5}$ pc, and we expect that $0 < \alpha < 1.5$.
In order to calculate the flux of gamma rays due to WIMP annihilation
near the Galactic center, we need to know the dark matter density
$\rho_0$ at radius $r_{\rm bh}=2$ pc.  This radius is at least two
orders of magnitude smaller than the best currently available
dissipationless simulations have reached.  Unfortunately, the
normalization of the dark matter profile in the central few pc is
quite uncertain, since it is affected by various phenomena associated
with the baryons which are the dominant mass component interior to the
solar radius, $d_{\sun}$.  A starting assumption is that $\rho_0$ is
given by extrapolating inward an NFW profile \cite{NFW:97}.  For
example, the fit of \cite{Stoehr:03}, corresponding to $\rho_{\rm
dm}(d_{\sun})= 0.46$ GeV$^2 c^{-4}$ cm$^{-3}$ with NFW scale radius
parameter $r_s=27$ kpc, implies $\rho_{\rm dm}(r_{\rm bh}) \approx 90
\ M_{\sun} \, \mbox{pc}^{-3}$.

Adiabatic compression of the dark matter due to baryonic infall
\cite{BFFP:86,Burkert:02} is likely to result in an increased central
density of dark matter.  For example, model A1 of the Galaxy in
\cite{KZS:02}, including the effects of baryonic compression, has
$\rho_{\rm dm}(100 \, {\rm pc}) = 10 \ M_{\sun} \, \mbox{pc}^{-3}$,
with the density increasing inward as $r^{-1.4}$; even if it scaled
from 100 pc to $r_{\rm bh}$ only as $r^{-1}$, this would give
$\rho_{\rm dm}(r_{\rm bh}) \approx 560 \, M_{\sun} \, \mbox{pc}^{-3}$.
On the other hand, in model B1 from \cite{KZS:02}, in which half of
the angular momentum of the baryons is assumed to be transferred to
the dark matter, $\rho_{\rm dm}(100 \, {\rm pc})$ is approximately 5
times lower than in model A1.  In addition, if the black hole formed
via mergers of the lower-mass black holes during the early stages of
galaxy formation, the density may be reduced further \cite{MMVJ:02}.
For definiteness, we take $\rho_0 = \rho_{\rm dm}(r_{\rm bh}) = 100 \,
M_{\sun} \, \mbox{pc}^{-3}$ as our fiducial value, although it is
likely that baryonic infall increased this.

The flux at the earth from a region of angular radius
$\sigma_\theta=0.05^{\circ}$, subtending solid angle $\Delta \Omega = \pi
\sigma_\theta^2$, centered on the black hole at the center of the
Galaxy is then 
$$
{\rm Flux} = {{N_\gamma \left<\sigma_{\rm ann} v\right> I}
                        \over {2 d_\odot^2 m_\chi^2}} \ .
$$
Here the number of gamma rays above the threshold energy $E_{th}$ of
the detector is approximately \cite{TO:02}
$$
N_\gamma = {5\over 6}x^{3/2} - {10\over3} x + 5x^{1/2} + 
   {5\over 6}x^{-1/2} - {10\over3} \ ,
$$
where $x\equiv E_{th}/m_\chi c^2$ and $m_\chi$ is the WIMP mass.
Assuming $E_{th}=50$ GeV and $m_\chi=100$ GeV $c^{-2}$, $x=0.5$ and
$N_\gamma = 0.0087$.
Taking the distance of the sun from the center of the Galaxy to be
$d_\odot = 8$ kpc \cite{Eis:03} and assuming $\alpha=1$ for definiteness,
\begin{eqnarray*}
{I \over {d_\odot^2}} &=& \int_L^{\sigma_\theta d_\odot} \rho_{\rm dm}^2 
    \left({r\over d_\odot}\right)^2 dr \\
&=& \rho_0^2 {{r_{\rm bh}^2} \over {d_\odot^2}} \left[ 
r_{\rm bh} {\ln {r_{\rm bh}\over L}} + (\sigma_\theta d_\odot 
- r_{\rm bh}) \right] \\
&=& 0.73 [24.4 + 5.2] {\rm pc} \, 
 \left({\rho_0 \over {100 M_\odot {\rm pc}^{-3} }}\right)^2
  {\rm GeV}^2 c^{-4} {\rm cm}^{-6}  \ .
\end{eqnarray*}

The minimum detectable WIMP annihilation cross section times velocity is
then
$$ 
\left<\sigma_{\rm ann} v\right>_{\rm min} =
    { {2 M_s m_\chi^2 N_{\rm bcg}^{1/2} } \over
      {N_\gamma A_{\rm eff} t I d_\odot^{-2} } } \ .
$$
For an ACT with $E_{\rm th}=50$ GeV, $N_{\rm bcg} = (3.7 + 7.9) \times
10^{-6}$ cm$^{-2}$ s$^{-1}$ sr$^{-1} \, A_{\rm eff} \, t \, \Delta
\Omega$, where the first term in the parenthesis is the
electron-induced background and the second is the hadronic background.
Since it is expected that the latter can be significantly reduced for
the new generation of ACTs \cite{Bergstrom:98,Stoehr:03}, we only
include the electron-induced background.  Then for a detection of
statistical significance $M_s \sigma$
(taking $M_s=3$ as our fiducial value),
\begin{eqnarray*}
\left<\sigma_{\rm ann} v\right>_{\rm min} & = &
   3.2 \times 10^{-26} \ {\rm cm}^3\ {\rm s}^{-1} \
   {M_s \over 3} \left({m_{\chi} \over 100 {\rm GeV} \ c^{-2} }\right)^2 \\
&& \left({0.0087\over N_\gamma}\right) 
   \left({A_{\rm eff} \over 10^8 {\rm cm}^2}\right)^{-1/2}
   \left({t \over 250 \ {\rm hr}}\right)^{-1/2}  \\
&&
   \left({\rho_0 \over 100 \, M_{\sun} \, {\rm pc}^{-3}}\right)^{-2} \ .
\end{eqnarray*}

The solid curve in Figure 1 shows this detection threshold as a
function of the WIMP mass $m_\chi$ for our fiducial assumptions.  The
dependence on $m_\chi$ here comes entirely from $m_\chi^2/N_\gamma$,
but a more realistic calculation would take into account the
increasing $A_{\rm eff}$ for higher gamma ray energy.  In addition,
the significance of the detection would be increased by including the
gamma rays from the full field of view of the ACT, which is
$5^{\circ}$ for the H.E.S.S. ACT array in Namibia \cite{HESS:03},
where the Galactic center passes nearly overhead.  The dashed curve
shows an order of magnitude improved sensitivity from baryonic
compression increasing $\rho_0$ by a modest factor of $10^{1/2}$.
Since at least that much baryonic compression is rather likely
\cite{Prada:04}, it is quite plausible that annihilation from the
Galactic center will be detected by new generation ACTs (H.E.S.S.,
CANGAROO III, MAGIC, and VERITAS) if the dark matter is actually WIMPs
of mass $m_{\chi} \gsim 100$ GeV $c^{-2}$.

\begin{figure}
\includegraphics[width=.5\textwidth]{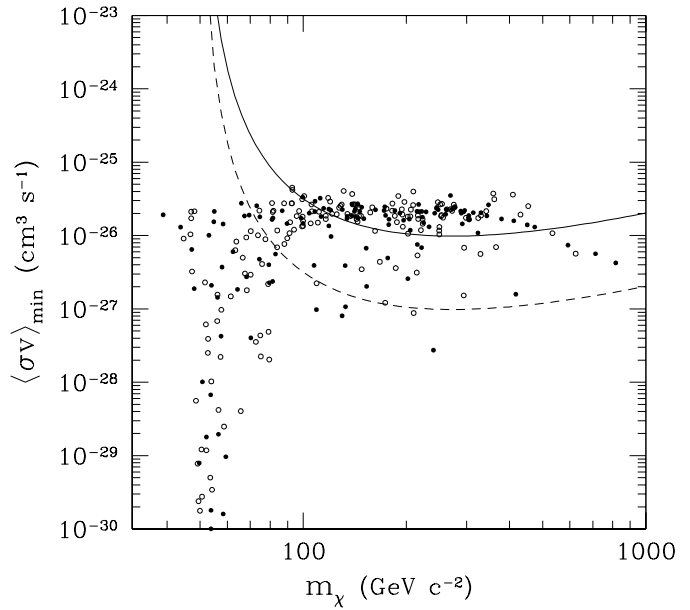}
\caption{
Minimum detectable annihilation cross section times velocity
as a function of WIMP mass.  The filled circles correspond to SUSY
model WIMPs with $\Omega_\chi h^2 = 0.11 \pm 0.01$
\protect\cite{Spergel:03} and the open circles correspond to SUSY
models with $\Omega_\chi h^2$ between 1$\sigma$ and 2$\sigma$ away from the
central value.}
\label{fig}
\end{figure}

The best strategy for detecting WIMPs of lower mass probably involves
using the all-sky map that will be produced by the Gamma Ray Large
Area Space Telescope (GLAST) satellite to look for the wide angle
annihilation flux away from the Galactic center, which can be reliably
calculated using dissipationless simulations \cite{Stoehr:03} since it
is hardly affected by uncertain physics such as baryonic compression.
The recent reanalysis \cite{HooperDingus:02} of the EGRET data toward
the Galactic center using a gamma ray energy dependent point spread
function already constrains some SUSY models with low-mass WIMPs,
subject to the uncertainties we have discussed about the dark matter
density toward the Galactic center.

{\it To summarize}, if the WIMP mass $m_\chi \gsim 100$ Gev $c^{-2}$
and the central density is only a little higher than our fiducial
value $\rho_0 = 100 \ M_{\sun} \, {\rm pc}^{-3}$ due to baryonic
compression, then the high angular resolution of the new ACTs such as
the H.E.S.S. array should permit detection of the $\rho_{\rm dm}
\propto r^{-3/2}$ central cusp that is inevitably associated with the
dense star cluster around the central black hole.


\bigskip
\begin{acknowledgments}
We would like to thank A. Kravtsov for organizing the workshop at the
Center for Cosmological Physics in Chicago in August of 2002 which
motivated this work.  OYG is supported by STScI Institute Fellowship.
JRP is supported by grants NASA-NAG5-12326 and NSF AST-0205944 at
UCSC, and he also thanks L. Stodolsky and S. White for hospitality in
Munich.  We enjoyed discussions with S. White, A. Klypin, F. Prada,
and F. Stoehr, and we especially thank Stoehr for giving us the output
\cite{Stoehr:03} from his run of DarkSUSY used in Figure 1.  We thank
the referee D. Merritt for comments that helped us improve the paper.
\end{acknowledgments}

\end{document}